# Deep Learning based Data Prefetching in CPU-GPU Unified Virtual Memory

Xinjian Long[a,b], Xiangyang Gong[a,b,*], Huiyang Zhou[c] and Bo Zhang[a,b]

[a]*State Key Laboratory of Networking and Switching Technology, Beijing University of Posts and Telecommunications, Beijing, 100876, China*
[b]*School of Computer Science (National Pilot Software Engineering School), Beijing University of Posts and Telecommunications, Beijing, 100876, China*
[c]*Department of Electrical and Computer Engineering at North Carolina State University, Raleigh, 27606, NC, USA*

## ABSTRACT

Unified Virtual Memory (UVM) relieves the developers from the onus of maintaining complex data structures and explicit data migration by enabling on-demand data movement between CPU memory and GPU memory. However, on-demand paging soon becomes a performance bottleneck of UVM due to the high latency caused by page table walks and data migration over interconnect. Prefetching is considered a promising solution to this problem given its ability to leverage the locality of program memory access patterns. However, existing locality-based prefetching schemes can not handle all the situations. An ideal prefetcher should not only look at narrow regions of the requested address space but also capture global context to deliver a good prediction of the memory access pattern.

This paper proposes a novel approach for page prefetching for UVM through deep learning. We first show that a powerful Transformer learning model can provide high accuracy for UVM page prefetching. We then perform analysis to interpret this Transformer model and derive several insights that allow us to design a simpler model to match the unconstrained model's accuracy with orders of magnitude lower cost. We evaluate this simplified model on a set of 11 memory-intensive benchmarks from popular benchmark suites. Our solution outperforms the state-of-the-art UVM framework, improving the performance by 10.89%, improving the device memory page hit rate by 16.98% (89.02% vs. 76.10% for prior art), and reducing the CPU-GPU interconnect traffic by 11.05%. According to our proposed unified metric, which combines the accuracy, coverage, and page hit rate, our solution is approaching the ideal prefetching scheme more than the state-of-the-art design (0.90 vs. 0.85, with the perfect prefetcher of 1.0).

## 1. Introduction

Modern GPUs support unified virtual addressing (UVA) [1, 2], which provides a unified virtual address space between the host CPU and the GPU. This technique allows the CPU to manage data residing in GPU physical memory with the same pointers as the ones used by the CPU program. Depending on UVA, UVM (Shared Virtual Memory or SVM in OpenCL terminology) supports the programmer-agnostic demand-driven movement of data between host and GPU memory instead of requiring the developers to manually copy data from the CPU to the GPU memory before a GPU kernel can access that data. This technique is typically supported by fault-driven transfers at the page or the multi-page (4KB to 2MB) granularity. Since CUDA 8.0, developers can decently exploit the UVM technique by using **cudaMallocManaged** instead of **cudaMalloc** followed by **cudaMemcpy** API calls in their programs.

However, along with the improved programmability, UVM also raises the consideration of the efficiency of the GPU runtime's (CUDA runtime in this paper) data management policy. As for UVM, only one physical copy of the data is maintained either on the host or the device memory. Following the first-touch migration policy, on every first access to a page by the device, the corresponding page table entry in the host/device is invalidated and data is migrated to the device/host memory and a new entry is created in the page table. On-demand paging will soon make UVM become costlier than the traditional **cudaMemcpy** due to a large amount of stalled warps and the near sequential combination of kernel execution and data migration. Prefetching is a good solution to these problems. But a poor prefetching scheme may easily generate heavy transfer traffic between the host and the GPU, which may also easily negate the potential performance benefits brought by the UVM technique.

Thanks to the attempts in recent studies [3, 4], researchers reveal that pages accessed by most of the GPU applications are strongly clustered. This locality mainly resides in virtual address space (physical addresses of the accessed pages may be non-contiguous due to address interleaving performed by the memory controller). To exploit this property, the locality-based data prefetching scheme has been introduced. This technique suggests that when the GPU runtime is provided with a requested page to be migrated, the runtime also schedules an additional $N$ pages in its virtual address neighborhood for migration. Locality-based prefetching has been proved effective in mitigating the overhead of using UVM. A prior work [5] also uncovered that a tree-based neighborhood hardware prefetcher was implemented in modern CUDA runtime.

However, locality-based prefetching techniques can not handle all the situations. When the applications preserve their page access pattern for certain time intervals or the memory access pattern is constant and repetitive, using locality-based prefetching will lead to significant performance improvement. However, when the subset of frequently accessed pages becomes disjoint between consecutive kernel iterations, application performance may drop drastically because the prefetcher fails to capture the locality from the latest memory accesses.

*Corresponding author
xygong@bupt.edu.cn (X. Gong)
ORCID(s): 0000-0002-0631-9747 (X. Gong)

Studies [3, 4, 5, 6, 7, 8, 9] have been proposed to improve locality-based prefetching. One of the simplest ideas is to increase the aggressiveness of the prefetcher. By retrieving more areas around the neighborhood of the first touched page, the chance of hitting the future requested pages also increases. However, this method also enlarges the probability of retrieving useless pages which would contribute little or nothing to the overall performance. Moreover, when the GPU device memory is oversubscribed, using an aggressive prefetching scheme may force the runtime to keep evicting the pages from the device memory, which causes a high risk of suffering page thrashing overhead. There are other more sophisticated methods like pinning the pages in the host memory and only access them remotely (i.e., hard pinning) or delaying the page migration with page-level access counters (i.e., soft pinning). These methods do not mitigate the prefetching scheme's strong reliance on locality, and they may also cause other problems like contention in the CPU-GPU interconnect and/or low utilization of powerful GPU local bandwidth.

This paper presents a novel approach for improving page prefetching in CPU-GPU UVM using deep learning. Our work has three steps. First, we design a powerful, unconstrained Transformer model that is trained for individual benchmark applications. Second, we interpret this model to reveal several important insights from the input data distribution and the output results. Third, we use these insights to simplify our design, and the revised design matches the unconstrained model's accuracy with orders of magnitude lower cost.

In choosing an unconstrained model, we build on the motivation that the accuracy of predicting memory access patterns should take long memory access history into consideration instead of blindly focusing on the neighborhood of the recently accessed region. Thus, we formulate the data prefetching problem as a sequence classification problem, where the goal is to assign each sequence of memory access history a category that indicates which page should be considered as a future prefetching candidate. We run different benchmark kernels on a GPGPU-Sim extension and collect the memory access traces. After some pre-processing, we use these traces as the training data and feed them to the Transformer model. We find that this model achieves high prediction accuracy across different GPU applications (Table 1).

An analysis of the Transformer-based model reveals several insights. First, clustering is a necessary step to help the predictor deliver good prediction performance, and the training data clustered by streaming multiprocessors (SM) id delivers the highest accuracy. Second, the Transformer-based model can adapt to different prediction distances while retaining high prediction accuracy. Third, the page address delta and the program counter (PC) make the major contribution to the high accuracy among several features that are extracted from the memory trace. Fourth, the full attention module is the Transformer-based predictor's main source of complexity. We can replace the full attention module with an approximation that matches the original model's prediction accuracy while reducing both the temporal and memory complexity. The level of approximation is determined by the distribution of the specific benchmark's memory access history.

We use these insights to design a simplified model, and we exploit the quantization technique to make its memory consumption be orders of magnitude lower than the unconstrained one (Table 6 and Table 7). The simplified model matches the unconstrained model's accuracy (Table 8) and it is more approaching to the theoretical upper bound of the prefetching scheme compared to the state-of-the-art design (Table 11).

To summarize, this paper makes the following contributions:

1. We present the first use of deep learning to page prefetching in CPU-GPU UVM.
2. We design a Transformer-based model that achieves very high prediction accuracy and can be interpreted to derive important insights about page prefetching. And we design a simplified attention module for this model to lower its complexity from $O(N^2)$ to $O((\log N)^2)$.
3. Our simplified model improves upon UVMSmart [9], a state-of-the-art framework for CPU-GPU UVM. Among 11 different GPU benchmark applications across different categories, our solution improves the benchmark IPC by 10.89% (geometric mean), improves the device memory page hit rate by 16.98% (89.02% vs. 76.10% for UVMSmart, arithmetic mean), and reduces the CPU-GPU interconnect usage by 11.05% (geometric mean). We also propose a metric to integrate the accuracy, the coverage, and the page hit rate. Based on this metric, we show that our solution is more approaching the perfect prefetching (0.90 vs. 0.85 for UVMSmart, and the perfect prefetcher is 1.0) than UVMSmart.

The remainder of this paper is organized as follows. Section 2 presents the background and motivation of this work. Section 3 discusses the related work of this paper. Section 4 describes the design of the unconstrained Transformer-based predictor. Section 5 derives the insights from the unconstrained model. Section 6 describes the design of the simplified predictor. Section 7 compares the results of our simplified model with the UVMSmart framework.

## 2. Background & Motivation

In this section, we review the general mechanics of on-demand paging in CPU-GPU UVM, the soft and hard pining, and the software and hardware prefetcher following the NVIDIA/CUDA terminology. It is worth noting that techniques mentioned in this section as well as our design described in the following sections are adaptable to the other GPU architectures besides NVIDIA. We then describe the performance bottleneck associated with GPU UVM page prefetching, and we discuss how we can exploit deep learning to alleviate the bottleneck.

### 2.1. On-demand Page Migration and Soft/Hard Pinning

CPU-GPU UVM provides a single virtual address space accessible from both CPU and GPU. Using CUDA, developers can apply UVM by calling the **cudaMallocManaged** API to allocate data that can be accessed by both host code and GPU kernels with a single shared pointer. The functionality of Unified Memory is enabled by on-demand memory allocation and fault-driven page migration. In modern GPUs, load/store instructions use virtual addresses. When a scheduled thread/warp in an SM (Streaming Multiprocessor) generates a device memory access with virtual addresses, such virtual addresses are translated to physical ones before accessing data in the GPU L1 cache. The load/store unit (LDST) of that SM performs a translation lookaside buffer (TLB) lookup to find whether the translation for the issued memory access is cached in TLB or not. A miss in the last level TLB will be relayed to GPU memory management unit (GMMU), which performs a page table walk for the requested page. If there is a hit in either the TLB lookup or the page table walk, the translation will be returned and the requested data will be accessed within the GPU memory hierarchy. This is shown as sequence (1) in Figure 1.

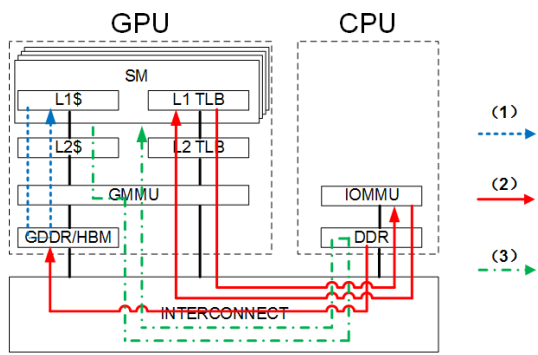

**Figure 1:** Overview of the UVM page migration and zero-copy.

However, if there is no page table entry (PTE) for the requested page or the valid flag is not set, then a far-fault is registered in the GMMU's Far-fault Miss Status Handling Registers (MSHR) and the corresponding warp will be stalled. Then this request will be forwarded to the host and triggered a host-side page table walk. Once the page table walk is finished and the requested page is returned, MSHRs will be consulted to notify the corresponding LDST to replay the device memory access, and then the stalled warp will be marked executable. This is shown as sequence (2) in Figure 1, and this is the general process of GPU UVM on-demand page migration.

Handling far-faults with on-demand migration is costly because of the high latency of page table walk and data migration over PCI-e interconnect. The NVIDIA CUDA runtime introduces pinning memory to alleviate this problem. On the one hand, developers can call the **cudaHostRegister** and the **cudaHostGetDevicePointer** APIs to force the memory allocation to be hard-pinned to the host memory. In this case, pages in such allocation of memory will never be transferred from host to device memory. GPU kernels can only request these pages using remote direct memory access (RDMA). This is demonstrated as sequence (3) in Figure 1, and this is the case of CUDA zero-copy. On the other hand, developers can call the **cudaMemAdviseSetAccessedBy** and the **cudaMemAdviseSetPreferredLocation** APIs to advise the allocation to be soft-pinned to the host memory. In this case, pages in such allocation will not be migrated to the device memory at the first touch. Rather, the migration will be delayed till the number of read-requests reaches a certain static threshold. This is illustrated as the combination of (2) and (3) in Figure 1.

### 2.2. Software and Hardware Prefetcher

CUDA 8.0 introduced the **cudaMemPrefetchAsync** API to handle the costly far-faults. This is a software prefetching scheme that allows the developers to manually overlap the kernel execution with the asynchronous data migration.

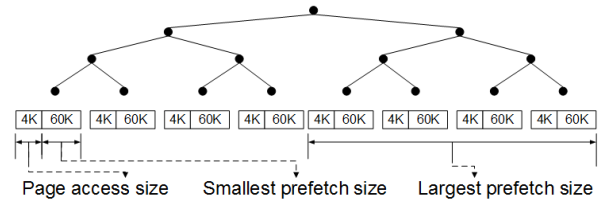

**Figure 2:** A tree-based neighborhood hardware prefetcher implemented by NVIDIA since CUDA 8.0 on a 512KB memory chunk.

In the GPU Technology Conference 2018, a tree-based hardware prefetcher was mentioned being implemented by NVIDIA CUDA 8.0 driver. Ganguly et al. [5] uncovered the semantic of this tree-based neighborhood prefetcher through micro-benchmarking and profiling. The user-requested size of a **cudaMallocManaged** allocation is logically divided into some 2MB memory chunks plus a remainder. Each of these chunk is further logically divided into 64KB basic blocks, which is the unit of prefetching. According to the far-faults received from the GPU, the runtime calculates the base addresses of the basic blocks corresponding to these faults. Then, these base addresses will be sent to the IOMMU and all the pages within the corresponding basic blocks will be migrated to the GPU. The runtime keeps track of the total size of valid memory resided in GPU for each non-leaf node among all the 2MB trees. If runtime detects that any non-leaf node's GPU valid memory is more than 50% of the total capacity of that node, the remaining non-valid pages of that node will be scheduled as further prefetching candidates. Figure 2 illustrates such a tree structure for a 512KB region.

### 2.3. Current Challenges

Studies [3, 4, 5] have shown that GPU applications' device memory accesses are strongly clustered in virtual address space. This further exemplifies the effectiveness of the tree-based prefetcher since it can prefetch at most $2MB$ neighborhood of the requested page. However, leveraging such (spatial) locality is not a panacea. Locality-based prefetching scheme performs well on GPU applications whose memory access pattern is constant and repetitive. When the subsets of hot pages become disjoint between consecutive kernel iterations , the locality-based prefetching scheme may soon become the performance bottleneck since it fails to capture the locality in the latest memory accesses and prefetches useless pages.

Most recently, Ganguly et al. [9] proposed an adaptive framework for oversubscription management in CPU-GPU UVM. This work achieves soft-pinning of hot pages to the device memory while remotely accessing cold pages from host memory by using adaptive runtime to detect pattern in CPU-GPU interconnect traffic. This work is promising but it does not solve the problem of locality-based prefetching scheme. On the other hand, remote zero-copy access may alleviate the large overhead caused by page prefetching. However, zero-copy requires very careful usage of pinned memory, since pinned memory buffers are limited and they are involved in both the pinned data transfer and the pageable data transfer. Excessive allocation of pinned memory may degrade the performance of both the host programs and the GPU kernels. Besides, relying on zero-copy transfer also leads to problems like low utilization of GPU memory bandwidth, as the interconnection bandwidth become new bottleneck.

In this work, we introduce deep learning to improve the page prefetching scheme. The Transformer architecture [10] and the Transformer-based models, such as BERT [11] and GPT-3 [12], are successful and yield state-of-the-art results on a wide variety of Natural Language Processing (NLP) tasks. This model is known to vastly outperform previous sequence model like LSTM. By extracting knowledge from a long GPU kernel memory access history instead of focusing on a narrow, currently requested range, we believe that deep learning will be a promising approach to improve the locality-based page prefetching scheme in CPU-GPU UVM.

## 3. Related Works

This paper is the first to propose deep learning based CPU-GPU UVM page prefetching scheme. We now discuss related works in UVM, and studies that apply artificial intelligence to other parts of the microarchitecture.

### 3.1. CPU-GPU UVM Studies

UVM support in modern discrete CPU-GPU systems [1, 2] has been studied widely. Agarwal et al. [3] proposed aggressive first-touch migration and prefetching neighboring pages. Zheng et al. [4] studied different user-directed and user-agnostic prefetchers to overlap data migration and kernel execution. Ganguly et al. [5] uncovered the mechanism of the tree-based prefetcher implemented in NVIDIA GPU driver. Pratheek et al. [13] proposed walk stealing to reduce interference in page walks from concurrent tenants while also ensuring high walker utilization. Chen at al. [6] proposed an application-transparent framework for reducing memory oversubscription overheads in GPUs. Kim et al. [7] proposed a GPU runtime software and hardware solution that enables efficient demanding paging for GPUs. Ganguly et al. proposed a programmer-agnostic framework [8] and an application-aware adaptive framework [9] to deal with memory oversubscription overhead stemming from page thrashing in irregular, data-intensive GPU applications.

### 3.2. Artificial Intelligence in Computer Architecture

Hashemi et al. [14] apply the RNN model into the analysis of memory access pattern, which demonstrates higher precision and recall than table-based approaches. Peled et al. [15] proposed the context-based prefetcher, which employs the contextual bandits model of reinforcement learning. Bhatia et al. [16] introduced perceptron-based prefetch filtering which acts as an independent check on the quality of predictions made by the underlying prefetch engine. Shi et al. [17] applied deep learning to solve the cache replacement problem. Doudali et al. [18] presented a page scheduler with machine intelligence for applications that execute over hybrid memory systems.

Peled et al. [19] use a fully-connected feed-forward network instead, and they formulate prefetching as a regression problem to train their neural network. Shi et al. [20] propose a hierarchical model of data prefetching that accommodates both delta patterns and address correlation. Bera et al. [21] propose a customizable prefetching framework that formulate prefetching as a reinforcement learning problem.

## 4. Transformer-based UVM Page Predictor

Inspired by the previous studies [14, 17], we formulate the prefetching problem as a classification problem. We phrase GPU UVM page prefetching as a supervised learning problem in which a predictor is trained with the past page accesses to predict the future page accesses. This is presented in Figure 3. Figure 4 shows the architecture of our predictor.

As shown in Figure 3, we collect the memory access traces of several GPU benchmark applications using an extended GPGPU-Sim [9], which supports UVM. We split each trace into a training and validation set, using 80% for training and 20% for validation, and we use 100% for testing. Instead of feeding the predictor with the entire benchmark trace, we cluster the data according to multiple features (demonstrated in Figure 3) and we try to find out the different levels of importance among them. This pre-processing exploration is further elaborated in Section 5.1.

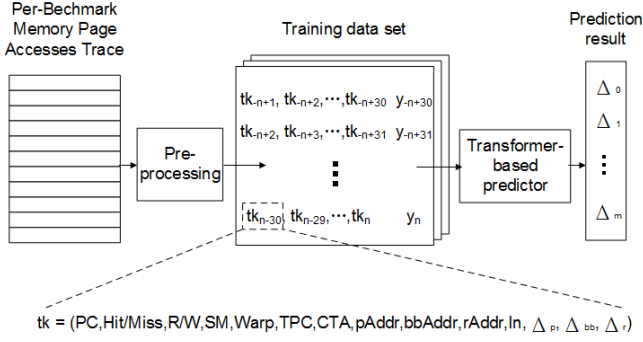

**Figure 3:** Overview of the Transformer-based UVM page prediction. A feature vector/token contains the following fields, PC: instruction address; Hit/Miss: the counter to tell whether this access causes a far-fault; warp/SM/TPC/CTA: the warp/Streaming Multiprocessor/Texture Processing Cluster/Cooperative Thread Array id; pAddr/bbAddr/rAddr: the 4KB page address/ the 64KB basic block address/the 2MB root node address of the requested page; In: the base addresses of the input arrays of the kernel function; $\Delta_p/\Delta_{bb}/\Delta_r$: the address delta of the corresponding type of addresses.

Hashemi et al. [14] found out that the number of uniquely occurring address deltas, $Addr(n) - Addr(n-1)$, is often orders of magnitude smaller than uniquely occurring addresses. To exploit this finding in our prediction, we label every unique delta within the trace, and we use them as the categories for classification. The number of categories varies among different benchmarks. Inspired by the experiments described in Section 7.2, we select Transformer as the basic model of this study. The Transformer-based predictor takes as the input a sequence of page accesses and assigns as the output a probability prediction to each classification category, where the prediction indicates the distance between the page addresses of the current and the next access.

Inspired by the prior work by Shi et al. [17], we consider that a long history of past accesses is beneficial and we define the sequence length as 30. We apply top-1 predictions on UVM page deltas. In our approach, for a faulty page, we keep prefetching its basic block, the same as the tree-based prefetching, and we will prefetch one additional page with the highest predicted probability to be accessed in the near future (i.e., top-1). This means that the maximum prefetching size of one read-request will be $15 + 1 = 16$ pages ($=64\,kB$), which will be much smaller than $1MB$.

In this work, we use an encoder-only Transformer, which is similar to BERT [11]. As shown in Figure 4, we begin by turning each token of the input sequence into a vector using embedding. The output size of the embedding layer is $200 * 30$ (30 is the length of the input sequence, and 200 is the total dimensions of the concatenation of 13 features mentioned in Figure 3 after embedding). After that, a positional embedding layer is applied and this is designed to help the predictor to recognize the token order. We use the original position encoding scheme [10], which is based on a family of sinusoidal functions. The output of the positional encoding layer will be input to a stack of Transformer encoders (we use the stack size of 2 in this work), which perform the multi-headed self-attention evaluation among the tokens within the input sequence and generate output encodings. Finally, these encodings are processed through linear transformation and softmax normalization, the prediction of the current input sequence will be generated. This finishes one forward pass of our predictor. The weights of this model are updated using back-propagation with gradient descent. Table 1 shows that the Transformer-based predictor achieves an average 94.04% top-1 accuracy and an average 0.9426 weighted f1 score among 9 benchmark applications, and these prove the effectiveness of the deep learning model for the page address prediction in CPU-GPU UVM.

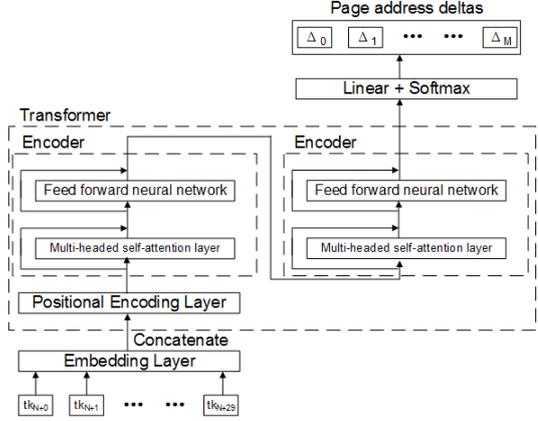

**Figure 4:** The Transformer-based predictor architecture.

**Table 1**
Transformer-based UVM page prediction results.

| Bechmark | f1 score | top-1 Acc. | top-10 Acc. |
|---|---|---|---|
| AddVectors | 0.9785 | 0.9767 | 0.9931 |
| ATAX | 0.9904 | 0.9943 | 0.9981 |
| Backprop | 0.9175 | 0.8893 | 0.9974 |
| BICG | 0.9932 | 0.9959 | 0.9992 |
| Hotspot | 0.7611 | 0.7676 | 0.9933 |
| MVT | 0.9889 | 0.9936 | 0.9979 |
| NW | 0.97 | 0.964 | 0.9958 |
| Pathfinder | 0.9128 | 0.9119 | 0.9996 |
| Srad-v2 | 0.9708 | 0.9707 | 0.9994 |

## 5. Insights from the Transformer Model

Although the Transformer-based predictor is effective in UVM page prediction, applying this model can be prohibitively costly in both time and memory. In this section, we discuss the insights observed from this predictor. We try to leverage these insights to improve the practicality of this model while retaining its effectiveness.

## 5.1. Preprocessing

Preprocessing is an effective method for the learning of memory access pattern [14]. By partitioning data into smaller clusters, the number of address deltas is significantly smaller than the global vocabulary, and this is beneficial for the model to extract knowledge from the traces. The PC is unique to an instruction that has been compiled from a program, and PC sequences can inform the model of patterns in the control flow. Researchers have used PC as one of the most important features [16, 17] when they try to extract knowledge from raw access traces. However, our observation (Table 2) shows that PC is not as effective as SM id when it is applied to clustering.

In GPGPU-Sim, we capture each benchmark kernel's memory trace from the GMMU. GPU tolerates long-latency stalls using fine-grained multithreading, and GPU cores may issue instructions from different warps to keep their pipeline busy. Each executing thread within these warps can access a different memory location. Thanks to the memory coalescing technique, the concurrent accessing requests will be alleviated to a smaller scale. However, there are still multiple page requests that may arrive at the GMMU simultaneously. These requests may come from different instructions depending on the SM processing speed, and they can be possibly mixed with each other losing the order information that the PC sequences are supposed to carry. We use the SM id to cluster the training data in this study.

**Table 2**
Page prediction results with different clustering methods.

| Bechmark | Cluster | f1 score | top-1 Acc. |
|---|---|---|---|
| AddVectors | PC | 0.4397 | 0.4403 |
| NW | PC | 0.6757 | 0.6029 |
| AddVectors | Kernel id | 0.4177 | 0.4168 |
| NW | Kernel id | 0.6948 | 0.603 |
| AddVectors | SM id | **0.9311** | **0.9152** |
| NW | SM id | **0.9562** | **0.9496** |
| AddVectors | CTA id | 0.4701 | 0.4588 |
| NW | CTA id | 0.7048 | 0.6207 |
| AddVectors | Warp id | 0.555 | 0.5001 |
| NW | Warp id | 0.7032 | 0.6207 |

## 5.2. Prefetching Timeliness

Timeliness is a critical factor of prefetching. Due to the high page migration latency through the CPU-GPU interconnect (PCI-e, NVLink, etc.), a too-early prefetch may never be able to reach the device memory before the associated far-fault occurs. On the other hand, a too-late prefetch is also unacceptable because the latency cost of demand access has already been paid before the prefetching is finished. Furthermore, when the memory is oversubscribed, a too-early prefetch may cause page thrashing which evicts the pages that have not been accessed yet and wastes the bandwidth of repeatedly transferring them. As shown in Table 3, our results indicate that the Transformer-based predictor can adapt to different prediction distances when it scales from 1 to 30. Empirically, we set the prediction distance as 30 in this study.

**Table 3**
Page prediction results using different prediction distances.

| Bechmark | Distance | f1 score | top-1 Acc. |
|---|---|---|---|
| Backprop | 1 | 0.9175 | 0.8893 |
| Srad-v2 | 1 | 0.9708 | 0.9707 |
| ATAX | 1 | 0.9904 | 0.9943 |
| NW | 1 | 0.97 | 0.964 |
| Backprop | 30 | 0.8895 | 0.7874 |
| Srad-v2 | 30 | 0.9444 | 0.9254 |
| ATAX | 30 | 0.9852 | 0.9888 |
| NW | 30 | 0.9801 | 0.9564 |

## 5.3. Input features

Among all the features described in Figure 3, we observe that 1) page address deltas, 2) page addresses, and 3) PCs carry most of the knowledge. It is worth noting that there are some special cases in this exploration. When we do similar tests on the data of ATAX, BICG and MVT, we found that no matter which feature is removed from the input sequences, the prediction performance difference is negligible. When we look into the data distribution, we find that there are always several address deltas dominant among the classification categories. For instance, as for ATAX, the number of training samples with address delta 16384 is 262077 while the training set size is 264040, which means address delta 16384 occupies 99.26% of the entire output vocabulary. In such cases, we will get almost the identical prediction results even we remove all the features in the input sequences. This finding inspires us on how to simplify the predictor according to the distribution of the input data.

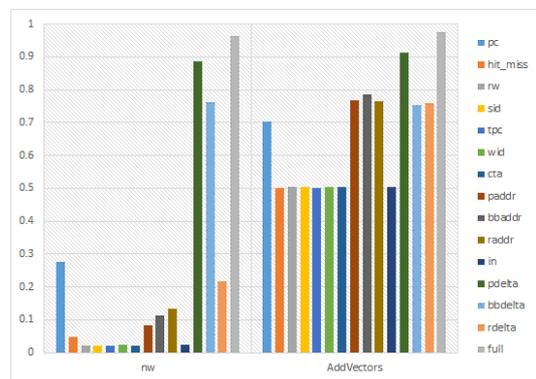

**Figure 5:** Page prediction results using one single feature. Features in this figure are the same as the ones demonstrated in Figure 3.

## 5.4. Full-attention module simplification

Simplification of the Transformer model is challenging and is a hot research topic recently [22, 23]. Researchers agree that the core limitation of the Transformer model is the quadratic dependency (mainly in terms of memory) on the sequence length due to the full attention mechanism. To improve the efficiency of the self-attention module, studies have been proposed to alleviate the heavy calculation brought by dot product. For example, CGNL [24] applies the Taylor series to approximate the pixel similarities. CC-Net [25] approximates the self-attention module via two consecutive criss-cross attention modules. These efforts give us the hint that we may not actually need a complete Transformer, and an approximation with less complexity may be good enough to solve our problem.

Thus, we do further explorations to find out the different levels of reliance on Transformer among all the benchmark kernels. We randomly shuffle the input sequences to see the effect of the self-attention module. It is worth noting that the original Transformer model is indifferent to the input sequence order. However, researchers argue that sequence order (word order) carries important information and it can make big difference semantically. We use the position encoder proposed in the original Transformer paper [10].

As demonstrated in Figure 6, ATAX, BICG, and MVT are not sensitive to input orders. These benchmarks have dominate address deltas as mentioned before. This indicates that the existence of dominant address delta not only weakens the contribution of other features, but also reduces the predictor's reliance on the self-attention module. We define the convergence of address delta as the ratio of the largest number of address delta to the total size of the output vocabulary. Figure 6 shows that the percentage of performance degradation of each benchmark without Transformer is proportional to its delta convergence. Thus, the convergence of address delta may be an important indicator to tell how many levels of self-attention module should be involved to perform accurate page address prediction. The remaining benchmarks show significant performance degradation when the input order are shuffled. And this indicates that the self-attention module plays an important role in the prediction of these benchmarks.

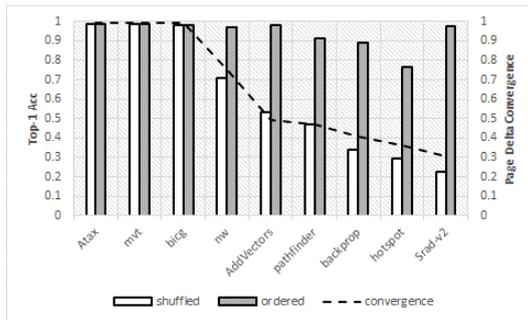

**Figure 6:** Page delta convergence and page prediction results for the ordered sequence and the shuffled sequence.

To verify our observation, we select four benchmarks in the previous exploration, two (ATAX, BICG) from the special cases and two (NW, Backprop) from the others, and we use Transformer and plain fully connected (fc) layers as the predictor separately. Table 4 shows that special cases work well even with one single fc layer without the self-attention module. In these cases, training using Transformer become totally unnecessary which only cause a waste of time and memory. The other two benchmark experience different levels of performance loss with only the fc layer, which indicates the necessity of Transformer in their predictions.

**Table 4**
Page prediction results using Transformer and simple fully-connected layer.

| Bechmark | Shuffle | Predictor | f1 score | top-1 Acc. |
|---|---|---|---|---|
| ATAX | True | Transformer | 0.9889 | 0.9939 |
| BICG | True | Transformer | 0.9932 | 0.9959 |
| NW | True | Transformer | 0.97 | 0.964 |
| Backprop | True | Transformer | 0.9175 | 0.8893 |
| ATAX | True | FC layer | 0.9894 | 0.9939 |
| BICG | True | FC layer | 0.9956 | 0.9929 |
| NW | True | FC layer | 0.8128 | 0.7378 |
| Backprop | True | FC layer | 0.4794 | 0.6666 |

We design our simplification method upon Reformer [22], a study that reduces the quadratic complexity $O(N^2)$ of the self-attention module to $O(N \log N)$ ($N$ indicates the length of the input sequence). The core idea of Reformer is to use locality sensitive hashing (LSH) attention, which is an approximation for full attention, to compute nearest neighbors and replace the $O(N^2)$ factor in attention layer, and enables the model to operate on long sequences. As described by Kitaev et al. [22], Reformer matches the results obtained with full Transformer but runs much faster, and has orders of magnitude better memory efficiency. Moreover, LSH attention is configurable, whose accuracy grows with the number of hashes. Therefore, we choose to leverage LSH attention in this work.

The key reason for the success of LSH attention is that the output of softmax normalization is dominated by the largest elements. So, when we perform the calculation of self-attention, which is $softmax(QK^T/\sqrt{d_k})$, we only need to focus on the $QK^T$ entries that may produce the largest dot product. In other words, we only need to focus on the $QK^T$ entries, which are closest to each other. By doing the calculation only on the closest entries instead of all the entries in the query and the key matrix, LSH attention can approximate the result of full attention with much less calculation overhead (both memory and time). Actually, this indicates an implicit precondition of using LSH attention. If the original attention matrix is not sparse enough, which means the $QK^T$ dot-product matrix delivers nearly uniform distribution, LSH attention may only be able to contribute little on reducing the complexity of the full attention.

Fortunately, when we look into the attention weights of our data, we find that the weight matrix is sparse enough even without configuring the scaling factor. As a result, we think that LSH attention is suitable for our problem.

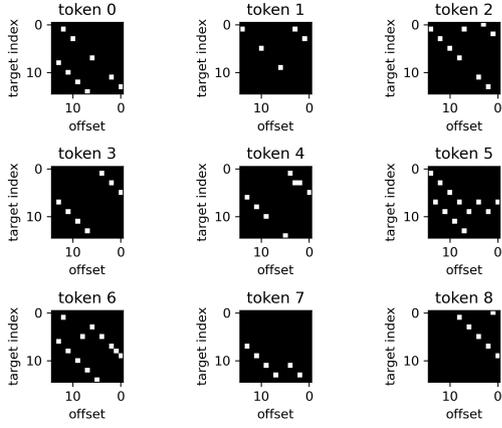

**Figure 7:** Attention weight vectors of consecutive tokens.

To further reduce the complexity of the attention mechanism, we propose Hamming-based Locality Sensitive Hashing Attention (HLSH) mechanism, which is described in Algorithm 1. In the original LSH Attention, after the sequences of queries and keys finish LSH bucketing and sorting, attention within each bucket will be calculated to generate the final output. We argue that it is not necessary to perform dot-product calculation among all the elements of each LSH bucket. Firstly, the results of dot-products are dominated by the similar $QK^T$ pairs. This indicates that the correctness of the final output will not be seriously affected even we erase the pairs whose distances are very large. Secondly, if vector *A* is very close to vector *B*, the dot-product *AC* will also be close to *BC*. We believe that similar queries should be able to share the dot-product results by running only one calculation while still providing a good approximation for the original attention matrix. As for each batch, we will get a matrix of hashes after LSH bucketing (as for the shared-qk structure, the LSH hashing of *Q* and *K* are identical). According to the definition of the angular LSH method described in Reformer, the more similar the input vectors are, the more likely the identical bucket ids will appear in the same round of angular LSH. Thus, we use hamming distance to measure the similarity of the LSH hashes, which gives the number of bucket ids that are different in the same bit (Line 2). As shown in in Algorithm 1, Line 3 is designed to reduce $seq\_len/2$ hamming values to one unified result for each entry. According to the aforementioned intuition, if the calculated hamming distance is larger than $HTOP$ (Line 6), we consider that the associated entry in the original *Q* or *K* matrix is very distinct compared to the other entries, which will lead to a negligible dot-product result in the final attention matrix. Thus, we simply erase such entries to save the time and memory of calculation (Line 7). On the other hand, if the calculated hamming distance is smaller than $HBOT$ (Line 9), this indicates that the associated *Q* entries are very close to each other, which will lead to a similar dot-product with the same *K* entry. In this case, we only keep the first entry and erase the others in this category, and we simply copy the dot-product result of the saved entry to the erased one according to their indexes (Line 9 − 16, Line 19). As for the entries associated with the calculated distance between $HBOT$ and $HTOP$, the typical matrix multiplication of attention will be operated. Assumed that the length of the LSH hashing is $L_{LSH}$, we define $HBOT$ as $0.1 * L_{LSH}$, and $HTOP$ as $0.9 * L_{LSH}$. Theoretically, our proposed HLSH attention mechanism can improve Reformer LSH attention from the complexity of $O(NlogN)$ to $O((logN)^2)$.

---

**Algorithm 1:** Hamming-based Locality Sensitive Hashing Attention Mechanism.

**Input:**
  The original shared matrix *Q* and *K*, whose shape are ($batch\_size, seq\_len, n\_embeds$).
  The resulting matrix of LSH bucketing $Q^{LSH}$ and $K^{LSH}$, whose shape are ($batch\_size, seq\_len, n\_hashes$).
  Threshold $HTOP$
  Threshold $HBOT$
**Output:**
  The resulting attention matrix $output^{HLSH}$.
1: **for** $i = 1 \to batch\_size$ **do**
2:   Randomly select $seq\_len/2$ entries from the sub-matrix $K_i^{LSH}$ according to $axis = 0$, and calculate their Hamming distance among all the entries in $Q_i^{LSH}$ according to $axis = 0$
3:   As for $Q_i^{LSH}$, calculate the *geomean* of the associated hamming distances according to $axis = 0$, get $HSCORE_i$
4:   Create buffer $RECORD$ whose max length is $seq\_len$
5:   **for** $j = 1 \to seq\_len$ **do**
6:     **if** $HSCORE_{ij} \geq HTOP$ **then**
7:       Replace the *j*th entry in $Q_i$ and $K_i$ with all-zeros vector
8:     **end if**
9:     **if** $HSCORE_{ij} \leq HBOT$ && *j* not in $RECORD$ **then**
10:      **if** the size of $RECORD$ is 0 **then**
11:        Record *j* as *base*
12:      **else**
13:        Replace the *j*th entry in $Q_i$ and $K_i$ with all-zeros vector
14:      **end if**
15:      Push *j* to $RECORD$
16:    **end if**
17:   **end for**
18:   $output_i^{HLSH} = matmul(Q_i, K_i^T)$
19:   Copy the entries in $output_i^{HLSH}$ whose indexes equals to *base* to the entries whose indexes equals to the values in $RECORD$
20: **end for**
21: Get $output^{HLSH}$

---

The comparison between using the full attention module and our proposed HLSH attention module is presented in Table 5. From the results, we can see that our proposed HLSH achieves very similar performance compared to the full attention module.

## 6. Our Solution

Our insights reveal that it is possible to substitute the Transformer-based predictor with a simpler model. We simplify and revise our predictor as follows:

1. We use the combination of SM id and warp id to cluster the traces.
2. We use 3 features (page address, page address delta, PC) to construct one input token, and we use 30 consecutive tokens to construct one input sequence. The output size of the embedding layer is decreased to 12 ∗ 30 (12 is the total dimensions of the concatenation of delta and PC after embedding) correspondingly.

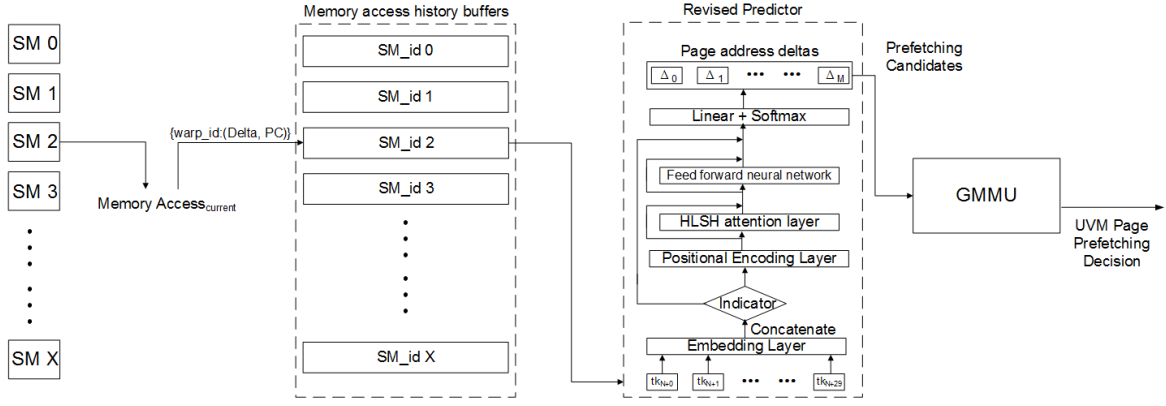

**Figure 8:** Overview of the revised predictor.

**Table 5**
Page prediction results using Transformer and HLSH attention.

| Bechmark | Shuffle | Predictor | f1 score | top-1 Acc. |
| --- | --- | --- | --- | --- |
| ATAX | True | Transformer | 0.9889 | 0.9939 |
| BICG | True | Transformer | 0.9932 | 0.9959 |
| NW | True | Transformer | 0.97 | 0.964 |
| Backprop | True | Transformer | 0.9175 | 0.8893 |
| ATAX | True | HLSH attention | 0.9887 | 0.9939 |
| BICG | True | HLSH attention | 0.9929 | 0.9956 |
| NW | True | HLSH attention | 0.95 | 0.9565 |
| Backprop | True | HLSH attention | 0.9209 | 0.8949 |

3. We use a single layer of encoder in the revised predictor.
4. We use HLSH attention to replace the full attention module, and the number of the attention head is 1.
5. We use 1 indicator to decide whether to bypass the attention module according to the page convergence of the input data.

The other parameters of the model are kept the same as the description in Section 4. Figure 8 shows the overall structure of the revised predictor.

Furthermore, we exploit quantization to compress the memory consumption of our model. Memory footprint of the previous model is shown in Table 6, which is nearly 187MB on average. Through our explorations, we find that clamping all the weights and the forward/backward pass activation values of all the layers to $[-8, +8]$ will not cause a serious performance degradation to the UVM page prediction. Table 8 shows the comparison results between using float32-based model and the clamped model. According to this finding, we believe that the memory consumption of the revised predictor could theoretically be one-eighth of the previous one (4 bits are enough to represent all the integers within $[-8, +8]$). The estimated memory consumption of the revised model is shown in Table 7, which is about 4.51MB on average. The memory consumption statistics are collected using the API provided by a MIT-licensed library [26]. (**torchinfo.summary**).

**Table 6**
Memory footprint using full-attention Transformer.

| Bechmark | Params. | F/B pass acti. | Total |
| --- | --- | --- | --- |
| AddVectors | 7.16MB | 151.10MB | 158.63MB |
| ATAX | 33.60MB | 151.44MB | 185.40MB |
| Backprop | 117.03MB | 155.33MB | 272.73MB |
| BICG | 27.87MB | 151.18MB | 179.41MB |
| Hotspot | 16.66MB | 151.27MB | 168.30MB |
| MVT | 33.48MB | 151.43MB | 185.28MB |
| NW | 39.27MB | 152.41MB | 192.05MB |
| Pathfinder | 25.96MB | 151.17MB | 177.50MB |
| Srad-v2 | 12.27MB | 151.22MB | 163.85MB |

**Table 7**
Memory footprint using revised predictor.

| Bechmark | Params. | F/B pass acti. | Total |
| --- | --- | --- | --- |
| AddVectors | 17.5KB | 4.30MB | ≈4.31MB |
| ATAX | 77.5KB | 4.34MB | ≈4.35MB |
| Backprop | 0.77MB | 4.83MB | ≈5.60MB |
| BICG | 31.25KB | 4.31MB | ≈4.32MB |
| Hotspot | 47.5KB | 4.32MB | ≈4.33MB |
| MVT | 76.25KB | 4.34MB | ≈4.35MB |
| NW | 0.25MB | 4.46MB | ≈4.71MB |
| Pathfinder | 28.75KB | 4.31MB | ≈4.32MB |
| Srad-v2 | 37.5KB | 4.31MB | ≈4.32MB |

## 7. Evaluation

We evaluate our simplified Transformer predictor by comparing it against a state-of-the-art framework (UVMSmart) which supports delayed page migration, zero-copy, and tree-based page prefetching. We compare IPC, page hit rate, CPU-GPU interconnect usage, and unity of benchmark kernels running with UVMSmart and our revised predictor. Unity, defined in Section 7.6, is our proposed single metric capturing the combined effects of prediction accuracy, coverage, and page hit rate on the GPU side.

**Table 8**
Page prediction results using Transformer(T) and revised predictor(R).

| Bechmark | f1 (T) | top1 (T) | f1 (R) | top1 (R) |
|---|---|---|---|---|
| AddVectors | 0.9785 | 0.9767 | 0.8986 | 0.9074 |
| ATAX | 0.9904 | 0.9943 | 0.9889 | 0.9939 |
| Backprop | 0.9175 | 0.8893 | 0.8267 | 0.8125 |
| BICG | 0.9932 | 0.9959 | 0.9929 | 0.9956 |
| Hotspot | 0.7611 | 0.7676 | 0.6846 | 0.7112 |
| MVT | 0.9889 | 0.9936 | 0.9886 | 0.9936 |
| NW | 0.97 | 0.964 | 0.8945 | 0.8821 |
| Pathfinder | 0.9128 | 0.9119 | 0.8734 | 0.8844 |
| Srad-v2 | 0.9708 | 0.9707 | 0.9335 | 0.9418 |

### 7.1. Evaluation methodology

We use an GPGPU-Sim extension implemented by Ganguly et al. [9] in our experiments. This extension provides functional and timing simulation support for UVM. Furthermore, this extension supports a smart runtime, which is composed of (1) a detection engine to identify the pattern in CPU-GPU interconnect traffic, (2) a dynamic policy engine that chooses from a wide array of existing memory management policies, and (3) an augmented memory management module that adaptively switches between delayed page migration and pinning. To compare with UVMSmart, we use the same GPGPU-Sim extension to run the same benchmark kernels with the same number of simulated instructions while disabling the UVMSmart runtime.

GPGPU-Sim UVMSmart provides a set of regular and irregular GPU applications from Rodinia, Lonestar, and Polybench benchmark suites. These benchmarks are modified to use CUDA UVM APIs. Since we focus on page prefetching in this work, we run these benchmarks under no oversubscription by configuring the device memory size larger than the benchmarks' working set size. Beside the benchmarks used in Sections 5 and 6, we use two more benchmarks (2DCONV, StreamTriad) in our evaluation. Table 9 shows the primary configuration of the simulator, and the configuration associated with the UVMSmart runtime is the same as in [9].

In order to hide the long latency of model training, we randomly select 5 benchmark applications (ATAX, Backprop, Bicg, Hotspot, NW) and run them using different input data set compared to the simulations described in Section 7. We use 50% of each of these benchmarks' simulation results to build a corpus, and we train our predictor described in Section 6 on this corpus until its accuracy reaches a reasonable range (≥0.85 in our experiments). We use this pre-trained model to make predictions for each benchmark, and we fine-tuned this model in each simulation every 50 million instructions to make it become adaptive in different program phases. According to our statistics among 11 benchmarks, this training method introduces a microsecond-level inference overhead for each prediction. According to NVIDIA's announcement [27], the inference latency of BERT-large (with 345 million parameters) could be slashed to 1.2 *ms* by leveraging the TensorRT 8.0 SDK. Shi et al. [20] also claim that their model (which is composed of LSTMs and Transformers with a much larger model size than our predictor) can make predictions every 18000 nanoseconds. We believe that this prediction overhead can be improved by more advanced hardware/software technologies, more fancy equipment, and more sophisticated programming skills. However, these are out of the scope of this study. Instead, we conduct a prediction overhead sensitivity test of our predictor (described in Section 7.3) to show the impact of predictor latency upon the performance.

**Table 9**
Configuration parameters of GPGPU-Sim.

| Simulator | GPGPU-Sim UVMSmart |
|---|---|
| GPU Architecture | NVIDIA GeForceGTX 1080Ti Pascal-like |
| GPU Cores | 28 SMs, 128 cores each @ 1481 MHz |
| Shader Core Config | Max 32 CTAs and 64 warps per SM, 32 threads per warp GTO scheduler |
| Page Size | 4KB |
| Page Table Walk Latency | 100 core cycles |
| CPU-GPU Interconnect | PCI-e 3.0 16x, 8 GTPS per channel per direction, 100 GPU core cycles latency |
| DRAM Latency | 100 GPU core cycles |
| Zero-copy Latency | 200 GPU core cycles |
| Far-fault Latency | 45$\mu$s |

### 7.2. Comparison among different predictors

As described in Section 3.2, researchers have adopted different AI-based approaches to divergent computer architectural problems. More precisely, some of these works [15, 16, 19, 20, 21, 28, 29, 30, 31] are targeting data prefetching problems (beyond GPU UVM), which is similar to our study. In order to compare these methods, we use a batch of 50 million instructions of each GPU application to train these models, and we use these trained models to make predictions on another batch of 50 million instructions. Figure 9 shows the comparison results of using different predictors to deliver page address prediction. We can see that the Transformer-based method delivers the best prediction performance compared to the other methods (Convolution Neural Network, LSTM, Multi-Layer Perceptron). And our revised predictor (HLSH) achieves a similar performance as Transformer.

It is worth noting that the input datasets of different GPU workloads are randomly generated, and our revised predictor achieves consistent performance on different GPU workloads' input datasets.

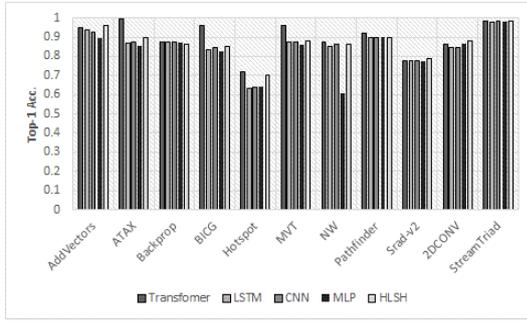

**Figure 9:** Prediction results using different predictors.

## 7.3. Prediction overhead sensitivity test

Figure 10 shows the normalized IPC results with different prediction latencies for 11 GPU benchmarks under no memory oversubscription. We vary the latency among 1, 2, 5, and 10 microsecond-per-prediction. Since the GPU core frequency is configured as 1481 MHz in the simulator, so these latencies roughly correspond to 1500, 3000, 7500, 15000 cycle-per-prediction in each simulation. We consider UVMSmart as the state-of-the-art (SOTA) design. The average normalized IPC results under different levels of prediction overheads are 1.10X (1 microsecond), 1.06X (2 microseconds), 1.00X (5 microseconds), and 0.90X (10 microseconds). Compared to the SOTA design, our predictor can achieve a 10% average IPC improvement when the prediction overhead is 1 microsecond, but this improvement vanishes when the overhead grows to 5 microseconds. Such deterioration continues and turns into a 10% performance slowdown when the overhead grows to 10 microseconds. These results show that our predictor, as well as other learning-based methods, are sensitive to the prediction overhead. In our subsequent experiments, we assume that our revised predictor is situated at the UVM backend to make predictions. We use 1 microsecond (1500 cycle) as the prediction overhead, which is sharply distinct from the previous works [17, 20] that consider zero prediction overhead while exploiting deep learning models to boost the application's IPC performance. Both the training overhead of the pre-trained model and the fine-tuning overhead are not considered in the simulation, we assume that these processes can be achieved offline in practice.

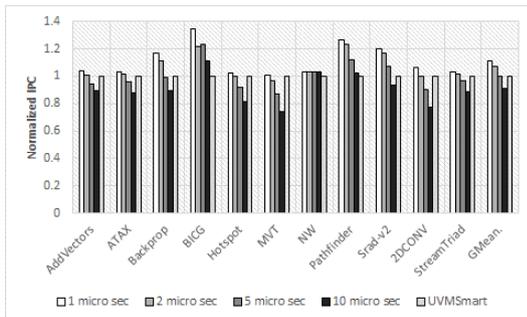

**Figure 10:** Normalized IPC of 11 GPU benchmark applications using our solution under different levels of prediction overhead.

## 7.4. Performance

From Figure 10, we can see that with the prediction latency of 1500 cycles, our solution achieves performance improvement for most benchmarks ranging from 1% to 34%. Compared to the aggressive tree-based prefetching method exploited by the UVMSmart runtime, our predictor improve the prefetching precision without harming the prefetching coverage (Table 11). By reducing the useless prefetches, page migration overheads caused by the corresponding load requests are saved. More accurate prediction also improves the application's page hit rate, which reduces the virtual address translation overhead. Srad-v2, Backprop's, and Pathfinder's IPC improvement is mainly due to this reason. Srad-v2's page hit rate grows from 0.86 to 0.94, Backprop's grows from 0.73 to 0.95, and Pathfinder's grows from 0.58 to 0.99 (Table 10). High-accuracy prefetching also helps mitigate the adverse effects of clustered page faults. For a far-fault, if multiple pages are to be transmitted as a result of prefetching, the service of the subsequent far-faults would be delayed since the PCIe bus has been occupied. As a result, some applications may experience high stall latency even the overall page hit rate is high. BICG follows this routine that the improvement in page hit rate is minor (from 0.94 to 0.99) while the IPC improvement is relatively significant (34%). This is further dissected in Section 7.5.

Overall, our solution improves IPC by an average of 10.89% (geometric mean) compared to UVMSmart. In addition, our solution improves the device memory page hit rate (i.e., the ratio of the demanded pages available at the GPU side) for all the benchmarks with an average (geometric mean) of 16.98%.

**Table 10**
Page hit rate(Hit) of GPU applications using UVMSmart runtime(U) and revised predictor(R).

| Bechmark | Hit(U) | Hit(R) | Simulated Inst. |
| --- | --- | --- | --- |
| AddVectors | 0.778014 | 0.940632 | 23068672 |
| ATAX | 0.979349 | 0.979211 | 13366000 |
| Backprop | 0.738671 | 0.955269 | 133000000 |
| BICG | 0.941975 | 0.998812 | 167936000 |
| Hotspot | 0.613115 | 0.840864 | 112000000 |
| MVT | 0.495362 | 0.509689 | 28000000 |
| NW | 0.999372 | 0.999874 | 29010432 |
| Pathfinder | 0.587640 | 0.994977 | 403828004 |
| Srad-v2 | 0.864029 | 0.942056 | 160000000 |
| StreamTriad | 0.559403 | 0.683084 | 7195000 |
| 2DCONV | 0.814476 | 0.948211 | 50212980 |

## 7.5. CPU-GPU interconnect usage

Figure 11 shows the PCIe usage of simulating BICG's 2 million instructions. As described in Section 2.2, the number of the prefetching pages grows when the valid pages of a 2MB node exceed 50%. We can see that UVMSmart's consumption of PCI-e bandwidth increases from nearly 1 GB/s to 15 GB/s following this scheme. More specially, the PCIe bus is highly occupied in the range of 407000 to 528000 cycles. In this case, the far-faults of the subsequent memory instructions have to wait until all the pending pages have been transferred. Thanks to the accurate page prediction powered by the revised predictor, our solution achieves higher prefetching precision without harming the prefetching coverage. As a result, the kernel using our solution experiences much less stalls compared to the one using the tree-based prefetcher. This is the reason why the simulation of the identical BICG's 2 million instructions using UVMSmart consumes 528244 cycles while our solution consumes 392440 cycles.

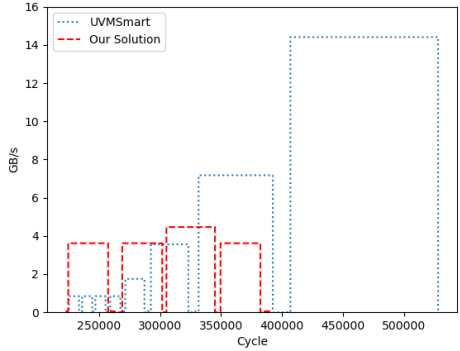

**Figure 11:** PCIe usage of BICG using UVMSmart runtime and our solution.

Overall, our solution reduces PCI-e usage by an average of 11.05% (geometric mean) compared to UVMSmart (Figure 12).

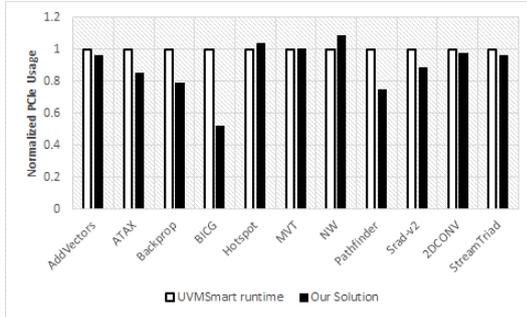

**Figure 12:** Normalized PCIe usage of 11 GPGPU benchmarks using UVMSmart runtime and our solution.

### 7.6. Unity

As described in previous works [16], prefetcher accuracy refers to the fraction of prefetched memory chunks that end up being used by the application, and prefetcher coverage refers to the fraction of memory access misses that could be mitigated by the prefetches. There is a fundamental tradeoff between prefetcher accuracy and prefetcher coverage. Also, another cirtical factor of prefetchers is the timeliness. To evaluate all these factors together, we propose a new metric 'unity', which is defined as follows.

$$Unity := \sqrt[3]{Accuracy * Coverage * Page\_hit\_rate} \quad (1)$$

As shown in the equation, we use page hit rate to measure the prefetching timeliness. For a perfect prefetcher, its unity would be 1. Next, we compare the unity of our proposed model with it of UVMSmart. The results are shown in Table 11. Our solution's average unity (0.90) is much closer to the ideal prefetching scheme (1.0) than UVMSmart (0.85). We can see that the UVMSmart's prefetching scheme (Tree-based Neighborhood Prefetcher) works really well at coverage (100% coverage among all the benchmark kernels). However, such perfectness comes at the price of a large number of prefetches, which affect the performance of the prefetcher accuracy (79.46% on average). Thanks to the deep learning model, our solution reduces the number of prefetches for most of the benchmark kernels without losing the prefetching correctness. As a result, our method achieves higher prefetcher accuracy (88.50%) while keeping a reasonable coverage (91.24%). By learning patterns in long memory access history instead of relying on most current accesses' spatial locality, our solution (89.02% on average) outperforms UVMSmart (76.10% on average) in the device memory page hit rate (i.e., timeliness). Both improved accuracy and timeliness are the reasons for the overall improved performance achieved by our proposed scheme.

**Table 11**
Unity of GPU applications using UVMSmart runtime(U) and revised predictor(R).

| Bechmark | Prefetcher | Acc. | Cov. | Hit. | Unity |
|---|---|---|---|---|---|
| AddVectors | U | 1 | 1 | 0.78 | 0.92 |
| ATAX | U | 0.89 | 1 | 0.98 | 0.96 |
| Backprop | U | 0.81 | 1 | 0.73 | 0.84 |
| BICG | U | 0.99 | 1 | 0.94 | 0.98 |
| Hotspot | U | 0.56 | 1 | 0.61 | 0.70 |
| MVT | U | 0.51 | 1 | 0.50 | 0.63 |
| NW | U | 0.99 | 1 | 0.99 | 0.99 |
| Pathfinder | U | 0.99 | 1 | 0.59 | 0.84 |
| Srad-v2 | U | 0.79 | 1 | 0.86 | 0.88 |
| StreamTriad | U | 0.51 | 1 | 0.56 | 0.66 |
| 2DCONV | U | 0.99 | 1 | 0.81 | 0.93 |
| AddVectors | R | 1 | 0.96 | 0.94 | 0.97 |
| ATAX | R | 0.90 | 0.88 | 0.98 | 0.92 |
| Backprop | R | 0.87 | 0.99 | 0.96 | 0.94 |
| BICG | R | 0.99 | 0.99 | 0.99 | 0.99 |
| Hotspot | R | 0.68 | 0.99 | 0.84 | 0.83 |
| MVT | R | 0.88 | 0.53 | 0.51 | 0.62 |
| NW | R | 0.99 | 0.98 | 0.99 | 0.99 |
| Pathfinder | R | 0.99 | 0.99 | 0.99 | 0.99 |
| Srad-v2 | R | 0.87 | 0.96 | 0.94 | 0.92 |
| StreamTriad | R | 0.68 | 0.92 | 0.68 | 0.75 |
| 2DCONV | R | 0.97 | 0.98 | 0.95 | 0.97 |
| | Ideal | 1 | 1 | 1 | 1 |

## 8. Conclusion

In this paper, we make a case for deep learning to improve page prefetching in CPU-GPU UVM. We first design a powerful Transformer-based model that uses the device memory access history of GPU applications, and this model delivers high prediction accuracy in the prediction of future requested pages. Then, we interpret the unconstrained Transformer model to derive several important insights, and we use these insights to simplify our model and to improve its practicality. The simplified model matches the high prediction accuracy of the unconstrained model with orders of magnitude lower cost. Finally, the evaluation results show that our proposed solution achieves higher performance than the state-of-the-art scheme for managing CPU-GPU UVM.

Thanks to the prior efforts of applying deep learning and machine learning algorithms into the area of microarchitecture, we can see a trend of exploiting machine intelligence to solve the valid problems (branch prediction, prefetching, cache replacement, etc.) in this domain. With the advance of technology, we expect that not only the simple models, such as perceptron, but also the more complex neural network, such as Transformer, could be leveraged by the hardware designers. We hope that this paper will inspire the design for other studies which are trying to introduce the learning-based solutions in heterogeneous systems like CPU-GPU and multi-GPUs.

## Acknowledgement

This work was supported in part by the National Natural Science Foundation of China (No.61802022 and No.61802027), and the Industrial Internet Innovation and Development Action Plan Project (No. TC210A02K).